\documentclass[twocolumn,prl,aps,showpacs,superscriptaddress]{revtex4}
\usepackage{epsfig} \usepackage{graphicx} \usepackage{epstopdf}

\newcommand{\be}{\begin{equation}} \newcommand{\ee}{\end{equation}}
\newcommand{\bea}{\begin{eqnarray}} \newcommand{\eea}{\end{eqnarray}}

 \newcommand{\void}[1]{}
\newcommand{\malpha}{a}
\newcommand{\tr}{\mathop{\mathrm{tr}}}

\begin{document}
\bibliographystyle{prsty}

\title{Non-Markovian Dissipative Semiclassical Dynamics}
\author{Werner Koch} \affiliation{Institut f\"ur Theoretische Physik,
  Technische Universit\"at Dresden, D-01062 Dresden, Germany}
\author{Frank Gro{\ss}mann} \affiliation{Institut f\"ur Theoretische
  Physik, Technische Universit\"at Dresden, D-01062 Dresden, Germany}
\author{J{\"u}rgen T. Stockburger} \affiliation{Institut f\"ur
  Theoretische Physik, Universit\"at Ulm, D-89069 Ulm, Germany}
\author{Joachim Ankerhold} \affiliation{Institut f\"ur Theoretische
  Physik, Universit\"at Ulm, D-89069 Ulm, Germany}

\begin{abstract}
  The exact stochastic decomposition of non-Markovian dissipative
  quantum dynamics is combined with the time-dependent semiclassical
  initial value formalism. It is shown that even in the challenging
  regime of moderate friction and low temperatures, where
  non-Markovian effects are substantial, this approach allows for the
  accurate description of dissipative dynamics in anharmonic
  potentials over many oscillation periods until thermalization is
  reached. The problem of convergence of the stochastic average at
  long times, which plagues full quantum mechanical implementations, is
  avoided through a joint sampling of the stochastic noise and the
  semiclassical phase space distribution.
\end{abstract}


\maketitle

\newpage The understanding of the nonequilibrium dynamics of open
quantum systems has been a central challenge in the last decades
\cite{Wei99,BP02}.  In recent years the subject has gained
considerable interest due to experimental progress which allows for
the tailoring and manipulation of quantum matter on ever larger
scales.  In mesoscopic physics, for instance, superconducting circuits
have been realized to observe coherent dynamics and entanglement
\cite{qubits}.  Similar advance has been achieved on molecular scales
with the detection of interferences in wave packet dynamics and the
control of the population of specific molecular states \cite{TdV02}.
Typically, these systems are in contact with a large number of
environmental degrees of freedom, e.~g.\ electromagnetic modes of the
circuitry or residual vibronic modes, which give rise to decoherence
and relaxation \cite{Zu91}. In many cases, the idealization of an
isolated system must inevitably be replaced by an open-system theory.

In the standard approach to open quantum systems, the reduced dynamics
of the system of interest is obtained by tracing out ``reservoir''
degrees of freedom from the conservative system-plus-reservoir
dynamics, e.~g., through projection operator techniques.
Alternatively, this program can be carried out through exact path
integral expressions for the reduced density matrix \cite{FV63}, which
became widely used in the 1980s \cite{Wei99}. The distinguishing
feature of dissipative path integrals is an influence functional which
describes self-interactions {\em non-local} in time.  Hence, a simple
quantum mechanical analogue to the classical Langevin equation is not
known; commonly used equations, such as Master/Redfield equations
\cite{BP02} in the weak-coupling case and quantum Smoluchowski
equations \cite{qse} for reservoir-dominated dynamics, rely on
perturbation theory. In intermediate domains, quantum Monte Carlo
techniques have been put forward for tight-binding systems, but the
achievable propagation times are severely limited by the dynamical
sign problem.

Recently, it has been shown that the influence functional can be
exactly reproduced through stochastic averaging of a process {\it
  without} explicit memory \cite{SG01,SG02}. The formulation turned
out to be particularly efficient for weak to moderate friction and low
temperatures \cite{SG01,St04}, a regime which lies beyond the strict
validity of Redfield equations on the one hand and beyond the
applicability of Monte Carlo schemes on the other hand.  The drawback
is though that for nonlinear systems the convergence of the stochastic
average for relatively long times is still an unsolved problem. Some
progress has been made for the spin-boson model by using a
hierarchical approach to quantum memory terms \cite{YYLS04}.  A
reliable and efficient, generally applicable method to tackle the
dissipative dynamics for continuous systems in this challenging
parameter regime is still missing.

Here we address this issue by combining the exact stochastic
Schr\"odinger formulation with a semiclassical real-time technique
based on a time-dependent initial value representation of the quantum
mechanical propagator. The central finding is that this procedure
circumvents the main obstacle of the exact formulation and allows for
accurate simulations up to times where equilibration sets in.  It is
important to note that a direct stationary-phase evaluation of the
double path integral for the reduced density is not a consistent
semiclassical approximation since the classical limit of open-system
dynamics, the Langevin equation, is not recovered \cite{Wei99,jcp95}.

Undamped systems allow a powerful method based on the semiclassical
propagator of Herman and Kluk (HK)\cite{HK84}, which has seen an
impressive number of applications ranging from atomic \cite{vdSR992}
to chemical physics \cite{TW04} after the work of Kay \cite{Kay941}
stimulated renewed interest in the approach. The HK
propagator was recently shown to be the leading term of consistent
series expansions of the {\em exact} quantum propagator of a Hamiltonian
system \cite{Pollak03,Kay06}. Currently, efforts are under way to directly
extend this approach to dynamics with quantum memory effects
\cite{Po07}. In this Letter, we will use a memory-free representation,
which accounts for non-Markovian dynamics entirely through
correlations of complex noise forces and thus suits numerical
applications.

We start with the standard decomposition of the total Hamiltonian
\begin{equation}
\hat{H}= \hat{H}_{S}+\hat{H}_{B}+\hat{H}_{I}\,
\end{equation}
 as a sum of a
system part, that for reasons of simplicity shall here depend on one
degree of freedom $x$, a bath part consisting of an infinity of
harmonic oscillators together with a bilinear interaction between
them.  In case of a factorized initial density with a bath residing in
thermal equilibrium at temperature $T$ one derives a path integral
expression for the time-evolved reduced density matrix of the form
\cite{Wei99}
\begin{eqnarray}
\label{eq:rhot}
\rho(x_f,x_f',t)&=&\int dx_i dx_i'
\rho(x_i,x_i',0)\int {\cal D}[x_1]{\cal D}[x_2]\nonumber
\\
&& \exp\left\{\frac{i}{\hbar}(S_{\rm S}[x_1]-S_{\rm S}[x_2])\right\}
F[x_1,x_2],
\end{eqnarray}
where the two real time paths $x_1$ and $x_2$ run in
time $t$ from $x_i$ and $x_i'$ to $x_f$ and $x_f'$, respectively. They
are coupled by the influence functional, which takes the form
$F[y,r]=\exp(-\Phi[y,r]/\hbar)$ with
\begin{eqnarray}
\label{influence}
\Phi[y,r] &=&\frac{1}{\hbar}\int_{0}^t\!\!\!\! du\int_{0}^{u}\!\!\!dv
y(u)[L'(u-v)y(v)+2i L''(u-v)r(v)]\nonumber
\\
& &+{i\mu}\int_{0}^tdu y(u)r(u)\, ,
\end{eqnarray}
where $y=x_1-x_2$, $r=(x_1+x_2)/2$ denote difference and sum paths,
respectively.  The complex valued friction kernel $L(t)=L'(t)+i
L''(t)$ is related to the force-force auto-correlation function of the
bath and completely determined by its spectral density $J(\omega)$ and
inverse temperature $\beta$. The static susceptibility denoted by
$\mu=-\int_o^\infty du L''(u)/(2\hbar)$ is a property of the
reservoir.

In \cite{SG02} it has been shown that a stochastic unraveling of the
forward and the backward paths leads to
\begin{eqnarray}
  \label{eq:densityprop}
  \lefteqn{\rho(x_f,x_f',t)=\int dx_i\int dx_i' \rho(x_i,x_i',0)}\nonumber\\
  && \hspace{1cm}\times M[K_{z_1}(x_f,t;x_i,0)(K_{z_2}(x_f',t;x_i',0))^\ast]\, ,
\end{eqnarray}
where $M$ denotes the average over noise realizations $z_{j}$ ($j$=1,2),
with the noise augmenting the system actions via
\begin{equation}
\label{eq:act}
S_{z_j}[x_j]=S_S[x_j]+\mu\int_0^t du\, x_j(u)^2+\int_0^tdu\, x_j(u)
z_j(u)
\end{equation}
in the path integral expressions of the respective propagators
$K_{z_j}$. This stochastic unraveling differs from a similar one by
Strunz et al.\ \cite{strunz} through the appearance of {\em two} noise
variables, allowing for the elimination of quantum memory effects.

Representing a general initial density operator through $
\hat{\rho}(t=0)=|\Psi_1\rangle\langle\Psi_2| $ (or through an ensemble
of such projectors) leads to two Schr\"odinger equations
\begin{eqnarray}
\label{eq:s1}
i\hbar{|\dot\Psi_1\rangle}&=&\left[H_{\rm S} -\xi(t)x+\frac{\mu}{2}x^2
  -\frac{\hbar}{2}\nu(t)x\right]|\Psi_1\rangle
\\
\label{eq:s2}
i\hbar{|\dot\Psi_2\rangle}&=&\left[H_{\rm S}
  -\xi^\ast(t)x+\frac{\mu}{2}x^2
  +\frac{\hbar}{2}\nu^\ast(t)x\right]|\Psi_2\rangle,
\end{eqnarray}
where
$\xi(t)=\frac{1}{2}[z_1(t)+z_2^\ast(t)]$ and
$\nu(t)=\frac{1}{\hbar}[z_1(t)-z_2^\ast(t)]$. 
The reduced density matrix (\ref{eq:rhot}) is reproduced
exactly by averaging $\hat\rho$ obtained from equations (\ref{eq:s1})
and (\ref{eq:s2}) when the correlations of $\xi$ and $\nu$ reproduce
the integral kernel of the influence functional:
$M[\xi(t)\xi(t')] = L'(t-t')$,
$M[\xi(t)\nu(t')] = (2i/\hbar) \Theta(t-t')L''(t-t')$, and
$M[\nu(t)\nu(t')] = 0$
($\Theta$ denotes the Heaviside step function).

The linear equations (\ref{eq:s1}) and (\ref{eq:s2}) are formally
exact in the sense that their Monte Carlo sampling will eventually
converge to yield Eq. (\ref{eq:rhot}). However, they are of limited
use for practical calculations since individual samples do not stay
normalized, which slows down convergence and makes a direct samping
impractical \cite{BP02,SG02}. This slowdown can be reduced \cite{St04}
by an exact mapping of the stochastic processes (\ref{eq:s1}) and
(\ref{eq:s2}) to a trace-conserving process (similar to a Girsanov
transform \cite{GG91}). Observing that only the last terms in the
square brackets of equations (\ref{eq:s1}) and (\ref{eq:s2}) lead to a
change of $\tr\hat\rho$, the first step of the transform consists of
subtracting an arbitrary ``reference trajectory'' $\bar{r}_u$ from $x$
in these terms \cite{footnote1}. The effect of this modification is
canceled exactly by a corresponding change in the probability measure,
which can be represented by the substitution \begin{equation}
\label{eq:chitilde} \xi\to \tilde{\xi} = \xi-\int_0^t du \chi(t-u)
\bar{r}_u\,, \end{equation} where $\chi(u)=-\Theta(u) L''(u)/2\hbar$
is the response function of the reservoir. Details of the transform
are given in Refs. \cite{SG02} and \cite{St04}.
With $\bar{r}_u=\langle \Psi_1|x|\Psi_2\rangle_u$, the diffusion of
$\tr\hat\rho$ is eliminated. However, this can lead to subtle
mathematical difficulties limiting the times for which numerical
simulations are stable \cite{St04}. There are two situations known to
be free of such instabilities, namely, linear systems and the
classical limit.  The idea is thus to combine the stochastic quantum
dynamics with a semiclassical propagation scheme based on the frozen
Gaussian approximation pioneered by Herman and Kluk \cite{HK84} and
Heller \cite{He81}.

The propagation of {\em individual samples} of the stochastic
processes (\ref{eq:s1}) and (\ref{eq:s2}) by the HK propagator differs
from the evolution of a closed system only through the addition of
simple potential terms, up to quadratic order, to the system
Hamiltonian. Hence the asymptotic convergence properties of the HK
propagator for closed systems \cite{marti06,Kay06,swart07} are
`inherited' by our stochastic samples.

The semiclassical HK propagator is given in terms of a phase space
integral as
\begin{eqnarray}
  \label{eq:hk}
  \lefteqn{K(x_f,t,x_i,0)= \int\frac{dp_idq_i}{2\pi\hbar}
    \langle x_f|g_\gamma(p_t,q_t)\rangle}\nonumber\\
  &&\hspace{1.5cm} \times R(p_i,q_i,t)
  {\rm e}^{i S(q_i,p_i,t)/\hbar}
  \langle g_\gamma(p_i,q_i)|x_i\rangle.
\end{eqnarray}
Here complex valued Gaussian wave packets $\langle
x|g_\gamma\rangle\sim\exp\{-\frac{\gamma}{2}(x-q)^2+\frac{i}{\hbar}p(x-q)\}$
of fixed width parameter $\gamma$ have been introduced, centered
around the initial phase space points $p_i,q_i$ and the time-evolved
phase space points $p_t,q_t$, respectively.  The pre-exponential
factor $R$ contains a complex valued combination of stability matrix
elements and the action reads as in (\ref{eq:act}) with the
replacement of the noise force described in equation (\ref{eq:chitilde}).
Obtaining the final density matrix $\rho(x_f,x_f',t)$ involves three
Monte Carlo integrations, two over the forward and backward phase spaces
of the semiclassical propagators and an additional one over the
noise trajectory distribution.

The classical trajectory entering equation (\ref{eq:hk}) is obtained
from the quasiclassical dynamics of the fixed-width Gaussians under
the transformed versions of equations (\ref{eq:s1}) and (\ref{eq:s2}).
The complex forces $\xi$ and $\nu$ do not extend the phase space to
complex numbers: The frozen Gaussians are, up to a trivial phase
factor, coherent states $|\alpha\rangle = e^{-|\alpha^2|/2}\,
e^{\alpha \hat{a}^\dagger} |0\rangle$ with
\begin{equation}
\alpha = \sqrt{\frac{\gamma}{2}}\left(q
+ \frac{i p}{\hbar\gamma}\right)\,.
\end{equation}
It is clear from this equation that complex values of $q$ and $p$ lead
only to states already described by a real-valued phase space. In
compact form, the classical equations of motion derived from
(\ref{eq:s1}) and (\ref{eq:s2}) read ($j$=1,2)
\begin{equation}
  \label{eq:eomclassical}
  \frac{d}{dt} {\alpha_j} = \sqrt{\frac{\gamma}{2}} \left( \frac{p_j}{m}
    - \frac{i}{\hbar\gamma}V'(q_j) + \frac{i}{\hbar\gamma} f_j
  \right)
\end{equation}
with $f_1 = \tilde\xi + \frac{\hbar}{2} \nu$ and $f_2 = \tilde\xi^* -
\frac{\hbar}{2} \nu^*$, to be solved for real $q_j$ and $p_j$. Taking
the limit $\hbar \to 0$, and integrating by parts in equation
(\ref{eq:chitilde}), the classical Langevin equation is indeed
recovered from equation (\ref{eq:eomclassical}).

In the semiclassical context, the reference trajectory is again
obtained by demanding that the $\nu$-dependent terms in equations
(\ref{eq:s1}) and (\ref{eq:s2}) do not change the trace of the
sample, which leads to the simple condition
\begin{equation}
  \label{eq:rusc}
  \bar{r}_u = (\alpha_1 + \alpha_2^*)/\sqrt{2\gamma}\,.
\end{equation}
This definition of $\bar{r}_u$ makes reference to a single pair of
semiclassical trajectories. It is therefore advantageous to merge the
integrations over the two HK phase spaces and the function space of
noise trajectories $\xi(t)$ and $\nu(t)$ into a joint Monte Carlo
sampling scheme where new initial phase space points and a new noise
trajectory are drawn for each sample \cite{footnote2}.

To demonstrate the approach, in the remainder we will concentrate on
the dynamics in a one-dimensional Morse potential
\begin{equation}
V(x)=\hbar D\,
\{1-\exp[-\malpha (x-x_0)/x_0]\}^2\, ,
\end{equation}
which may serve e.~g.\ as a simple model for the relative motion of a
diatomic molecule with equilibrium separation $x_0$.  The HK
propagation for the isolated system is known to accurately reproduce
the exact quantum dynamics.  Here we introduce the scaled displacement
$\tilde{x}=(x-x_0)/x_0$ and the frequency unit $\hbar/(m x_0^2)$.  In
the following we use the parameters $D=30$ and $\malpha=0.08$
corresponding to a total number of 97 bound states.  In these units
the frequency $\omega_0$ for small oscillations around the well
minimum located at $\tilde{x}=0$ is $\omega_0=\malpha\sqrt{2 D}=0.62$.
In the Markovian case and with the restriction to the bound state part
of the spectrum a similar system has been studied in \cite{BHP97}.  As
the initial state of the propagation we have chosen a Gaussian wave
packet shifted away from the potential minimum with
$\langle\tilde{x}\rangle_0=1$ and zero initial momentum.  The spectral
density of the bath oscillators is taken to be Ohmic with an algebraic
cutoff
\begin{equation}
J(\omega)=\frac{\eta\, \omega}{(1+\omega^2/\omega_c^2)^2},
\end{equation}
where
the cutoff frequency $\omega_c=10$ is well beyond any relevant system
frequencies and the dimensionless coupling strength is denoted by
$\eta$.

\begin{figure}
  \includegraphics[width=8.2cm]{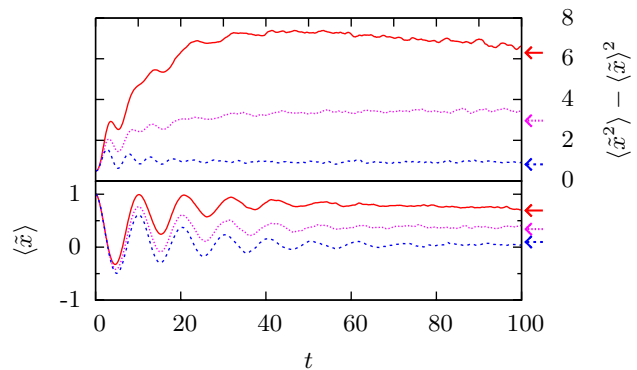}
  \caption{\label{fig:beta}Expectation value of position and variance
    in a dissipative Morse oscillator for decreasing temperature
    $\hbar\beta=0.5$ (full), $\hbar\beta=1$ (dotted), $\hbar\beta=10$ (dashed) at
    constant $\eta=0.1$. Arrows indicate thermal expectation values of
    the unperturbed oscillator (see text).}
\end{figure}
Let us first look at the temperature dependence of mean and variance
in position for a fixed interaction strength $\eta=0.1$, displayed in
Fig.~\ref{fig:beta}. As expected, damped oscillations occur for
intermediate times. In sharp contrast to the full quantum mechanical
case, however, the simulations are stable over long periods of time
until equilibration is approached. Arrows indicate equilibrium values
obtained by Boltzmann weighting analytic bound-state results from
\cite{Ga80}. At higher temperatures the mean saturates at a value
considerably away from zero thus revealing the substantial influence
of the anharmonicity. This influence gradually decreases when the
temperature is lowered. Only at the lowest temperature
($\hbar\beta=10$, dashed arrows) are the results consistent with a
harmonic approximation of the oscillator. Similar effects are seen in
the position variance, with the additional feature that coherent
oscillations are strongly smeared out at higher temperatures.
Importantly, for the lowest temperatures considered here, one has
$\omega_c\hbar\beta\gg 1$ so that retardation effects are strong and
the dynamics is far from being Markovian.  Further, since
$\eta\hbar\beta\approx 1$ we are outside the weak coupling regime in
the sense of Master/Redfield equations.  To check the thermal nature
of the final expectation values, we used different initial
preparations and found that the results converge to the same long time
average (not shown). The total number of sampling points needed to
generate all presented results is 10$^7$; the calculations take a few
hours on a desktop PC.

\begin{figure}
  \includegraphics[width=8.2cm]{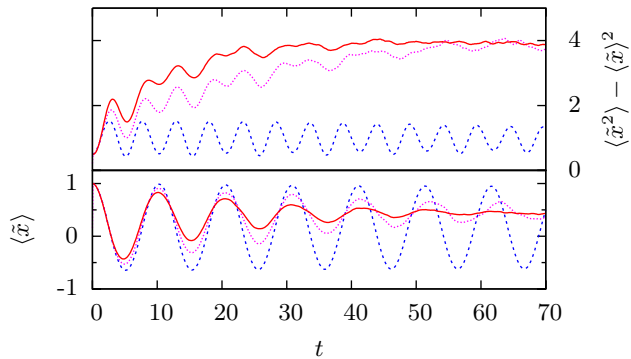}
  \caption{\label{fig:eta}Same as in Fig.~\ref{fig:beta}, but for
    increasing coupling $\eta=0$ (dashed), $\eta=0.05$ (dotted),
    $\eta=0.1$ (full), all at $\hbar\beta=1$.}
\end{figure}

In Figure \ref{fig:eta} the same expectation values are depicted at a
fixed inverse temperature $\hbar\beta=1$ but varying dissipation
strengths. Of course, pronounced coherent oscillations are seen for
vanishing coupling, where again the anharmonicity of the potential
shows up in smaller oscillation amplitudes for negative than for
positive values of position. Now a marked difference between the
isolated system and the case of weak damping can be observed: Even for
weak damping the amplitudes of the oscillations decrease strongly,
with centers shifted towards the soft side of the potential. The
variance increases as a function of time, while its initial
oscillations decay on a somewhat shorter timescale. We mention that at
the fixed temperature $\hbar\beta=1$, almost identical long time
averages are reached for all small friction strengths that we have
investigated.

To summarize, we have introduced a well-defined semiclassical
propagator for non-Markovian open quantum systems. It was derived by
consistently applying a known stochastic decomposition of the
influence functional appearing in exact open-system path integrals. A
test using a standard model of molecular physics shows that the
dissipative semiclassical propagator accurately describes the time
evolution over long periods of time, up to thermal equilibration
between open system and environment. With modest computational
resources, we provide data for this model which are difficult if not
impossible to obtain by other means in the low-temperature regime of
non-Markovian dynamics. For systems where corrections to the HK
propagator are needed, extensions according to \cite{Kay06} are
straightforward. We estimate that the number of samples needed for a
higher-dimensional quantum system will not increase substantially
beyond the number of samples in the one-dimensional case because the
phase-space sampling converges more rapidly than the noise average.

W.~K., F.~G., and J.~T.~S.\ acknowledge financial support by the
Deutsche Forschungsgemeinschaft (GR 1210/4-1, SFB 382) and J.~A. by
the German-Israeli foundation.



\begin{thebibliography}{10}
\bibitem{footnote1} The reference trajectory may be any causal,
  analytic functional of $\xi$ and $\nu$.
\bibitem{footnote2} Since the initial HK phase space density and the
  noise probability are independent, other sampling schemes are
  possible, but the close relation of equation (\ref{eq:eomclassical})
  to the classical Langevin equation may be lost.
\end{thebibliography}

\end{document}